# Observation of two-dimensional defect surface solitons

A. Szameit[1], Y. V. Kartashov[2], M. Heinrich[1], F. Dreisow[1], T. Pertsch[1], S. Nolte[1], A. Tünnermann[1,3], F. Lederer[4], V. A. Vysloukh[2], and L. Torner[2]

[1]*Institute of Applied Physics, Friedrich-Schiller-University Jena, Max-Wien-Platz 1, 07743 Jena, Germany*

[2]*ICFO-Institut de Ciencies Fotoniques, and Universitat Politecnica de Catalunya, Mediterranean Technology Park, 08860 Castelldefels (Barcelona), Spain*

[3]*Fraunhofer Institute for Applied Optics and Precision Engineering, Albert-Einstein-Strasse 7, 07745 Jena, Germany*

[4]*Institute for Condensed Matter Theory and Optics, Friedrich-Schiller-University Jena, Max-Wien-Platz 1, 07743 Jena, Germany*

We report on the experimental observation of two-dimensional solitons located in a defect channel at the surface of a hexagonal waveguide array. The threshold power for the excitation of defect surface solitons existing due to total internal reflection grows with decrease of the refractive index in negative defects and vanishes for sufficiently strong positive defects. Negative defects can also support linear surface modes existing due to Bragg-type reflections.

Discrete solitons [1,2] may emerge at interfaces of materials yielding different physical properties. Discrete surface solitons introduced recently for the interface of uniform and periodic nonlinear media [3,4] are particularly interesting since they combine the characteristics of surface states at uniform interfaces and the unique properties of lattice solitons arising from the periodicity of underlying medium. Surface solitons may exist not only at focusing but also at defocusing lattice interfaces [5-7]. Very recently the properties of two-dimensional surface solitons were analyzed theoretically [8-13] and experimentally [14-16]. In all these settings the discrete surface states were realized at the edge of perfectly periodic lattices. However, the addition of defects may substantially modify the guiding properties [17]. Lattices with positive or negative defects allow for linear localized states existing due to total internal or Bragg-type reflection [18]. When placed at the edge of a one-dimensional lattice, the defects strongly alter the properties of surface solitons [19,20]. Nevertheless, in two-dimensional structures the formation of nonlinear defect surface states and the generation of thresholdless surface waves were not observed.

In this Letter we show the existence of different types of two-dimensional defect surface solitons, relying on two different guiding mechanisms, and report on their experimental observation in femtosecond laser written waveguide arrays with hexagonal symmetry. We show that the threshold power and the excitation dynamics for surface solitons strongly depend on the type and strength of the surface defect.

In the simulations we employ the nonlinear Schrödinger equation for the dimensionless field amplitude $q$ under the assumption of cw illumination:

$$i\frac{\partial q}{\partial \xi} = -\frac{1}{2}\left(\frac{\partial^2 q}{\partial \eta^2} + \frac{\partial^2 q}{\partial \zeta^2}\right) - q|q|^2 - pR(\eta,\zeta)q. \qquad (1)$$

Here $\xi$ is the longitudinal and $\eta,\zeta$ are the transverse coordinates. The parameter $p$ stands for the refractive index modulation depth, while the function $R(\eta,\zeta)$ describes the refractive index profile in the array as a superposition of Gaussian functions $\exp[-(\eta/w_\eta)^2-(\zeta/w_\zeta)^2]$ modeling the individual waveguides, which are arranged in a hexagonal array with spacing $w_s$. The defect waveguide is located in the first row. The refractive index in the defect waveguide is given by $p=p_d$, while in other waveguides it is $p=p_a$. For positive defects one has $p_d > p_a$ and for negative ones $p_d < p_a$. We set $w_\eta = 0.45$, $w_\zeta = 0.9$, and $w_s = 4$ in accordance with the transverse waveguide dimensions of $4.5 \times 9$ $\mu m^2$ and the spacing of $40$ $\mu m$. We fixed $p_a = 2.8$ that corresponds to a real modulation depth $\sim 3 \times 10^{-4}$. In the simulations we model the array with a finite number of waveguides (according to the experiment) but results hold true for semi-infinite arrays.

The stationary profiles of surface solitons residing in the defect channel are of the form $q = w(\eta,\zeta)\exp(ib\xi)$, where $b$ is the propagation constant. Such solitons exist for $b$ values above a cutoff $b_{co}$ and they can be characterized by their total power $U = \int\int_{-\infty}^{\infty} w^2 d\eta d\zeta$. We found two types of defect surface states: those mediated by total internal reflection and those emerging from Bragg reflection. The states forming due to total internal reflection never change their sign between the waveguides [Figs. 1(a)-1(c)]. Their properties are sum-marized in Figs. 2(a)-2(c). When $p_\mathrm{d} \leq 2.94$ the $U(b)$ dependence for surface solitons is nonmonotonic [Fig. 2(a), curve 1]. Solitons exist only above the power threshold $U_\mathrm{th}$, which is also typical for surface solitons in defect-free arrays. The slope of $U(b)$ curve tends to infinity in the cutoff, where the soliton is most extended. Also, the threshold $U_\mathrm{th}$ increases with decreasing $p_\mathrm{d}$ [Fig. 2(b)], while the cutoff changes only slightly. In contrast, for $p_\mathrm{d} > 2.94$ the cutoff changes rapidly [Fig. 2(c)] while the threshold $U_\mathrm{th}$ vanishes [Fig. 2(a), curves 2 and 3], since strong positive surface defects support linear modes [shapes of low-power solitons close to linear modes are shown in Figs. 1(a) and 1(c)]. Note that for $p_\mathrm{d} \approx 2.94$, when the penetration of the field into the lattice is maximal, increasing or decreasing the defect strength results in a dramatic reduction of the light field penetration, even close to the cutoff. Far from the cutoff the light gradually concentrates in the defect site [Fig. 1(b)]. Since only surface states on the $dU/db > 0$ branches of the $U(b)$ curve are linearly stable, there exist narrow instability domains close to the cutoff for $2.94 < p_\mathrm{d} < 2.99$. For $p_\mathrm{d} > 2.99$ the surface solitons become stable in the entire domain of existence.

Negative defects ($p_\mathrm{d} < 2.8$) may support a second type of surface modes mediated by Bragg-type reflection from the periodic structure. The field in such modes exhibits a complicated staggered structure and changes its sign [Fig. 1(d)]. Nonlinear surface modes supported by negative defects thus bifurcate from such linear surface modes; hence their power vanishes in the cutoff [Fig. 2(d)]. Therefore, such states can be dynamically excited with very small powers. This is in sharp contrast to defect surface solitons mediated by total internal reflection which feature considerable thresholds at such small $p_\mathrm{d}$ values. The degree of localization of staggered surface modes increases and their narrow domain of existence expands with decreasing $p_\mathrm{d}$.

For the experimental observation of defect surface solitons we used waveguide arrays in fused silica glass fabricated with a femtosecond direct writing technique [21]. The length of our samples was 105 mm, while the spacing of the guides was 40 µm. The writing velocity was 1750 µm/s to keep the uniform nonlinear refractive index of the unprocessed material in the waveguides [22]. The defect sites were written with 1500 µm/s and 2000 µm/s for a positive defect and negative defect, respectively. For the excitation of the solitons, we used a 1 kHz fs laser system (Spectra Tsunami/Spitfire) with a pulse length of about 180 fs at a central wavelength of 800 nm. To confirm the experimental data by simulations, we solved

Eq. (1) with the input conditions $q|_{\xi=0} = A\exp[-(\eta - \eta_d)^2 - (\zeta - \zeta_d)^2]$, where $\eta_d, \zeta_d$ are the coordinates of the center of the defect surface waveguide. Note that our experiments are conducted with pulsed pump light, while the numerical simulations assume CW illumination; thus the comparison serves only to confirm the consistency of observations with soliton formation around the pulse peak.

In the array without defect (Fig. 3) the formation of nonlinear surface waves is similar to that in recent experiments [23]. At low powers one observes strong linear diffraction [Fig. 3(a)]. For intermediate power levels, the light gradually contracts towards the excited surface channel [Fig. 3(b)]. The formation of surface soliton is observed above the threshold power [Fig. 3(c)]. The picture changes dramatically in the case of a positive surface defect. Stronger defects result in a gradual suppression of light diffraction even at low powers and, eventually, in the formation of a linear surface mode. This is accompanied by a strong reduction of the power threshold for soliton excitation. Thus, an array with a strong positive defect ($p_d = 3.03$) supports a well localized linear mode shown in Fig. 4(a). At all power levels the light remains confined to the surface defect waveguide. An increase of the input power does not result in considerable changes of the mode profiles [see Figs. 4(b) and 4(c)].

We also observed light localization at low powers in the case of a negative surface defect for $p_d = 2.7$ [Fig. 4(d)], a result that we attribute to the existence of linear defect surface modes caused by Bragg-type reflection from the periodic structure. Notice that in the absence of defects surface modes, soliton excitation always occurs above a threshold power, for both total internal or Bragg reflection mechanisms of localization. Increasing the input power results in a matching of the total refractive index in the defect waveguide and in the array, which is accompanied by a sudden delocalization at intermediate power levels [Fig. 4(e)]. When the input power is further increased one excites the usual surface soliton existing due to total internal reflection [Fig. 4(f)]. The threshold power for the excitation of such solitons is much higher than in the case of defect-free arrays or positive surface defects.

In conclusion, we reported on the existence of two-dimensional defect surface solitons for both positive and negative defects. The inclusion of positive defects allows for the formation of surface solitons existing due to total internal reflection at a reduced threshold power. Negative defects may support surface states mediated by both, total internal reflection and Bragg-type reflection from the periodic structure.

# References with titles

# References without titles

**Figure captions**

Figure 1. Field distributions for defect surface solitons at (a) $b = 0.4644$, $p_{\rm d} = 2.96$, (b) $b = 0.512$, $p_{\rm d} = 2.96$, (c) $b = 0.49$, $p_{\rm d} = 3.1$, (d) $b = 0.274$, $p_{\rm d} = 1.7$. The white dashed lines indicate the interface position.

Figure 2. (a) Power versus propagation constant for $p_{\rm d} = 2.92$ (curve 1), $2.96$ (curve 2), and $3$ (curve 3). The circles correspond to solitons in Figs. 1(a) and 1(b). Threshold power (b) and cutoff (c) versus $p_{\rm d}$. Dashed lines indicate the point $p_{\rm d} = 2.8$. (d) Power versus propagation constant for a surface mode existing due to Bragg guiding at $p_{\rm d} = 1.7$. The circle corresponds to profile from Fig. 1(d).

Figure 3. Comparison of the output intensity distributions for an excitation of a surface waveguide in array without defect at $p_{\rm d} = p_{\rm a} = 2.8$. Top row - experiment, bottom row - theory. The input peak power is (a) $0.18$ MW, (b) $1.3$ MW, and (c) $3.2$ MW.

Figure 4. Output intensity distributions for an excitation of (a)-(c) positive surface defect waveguide at $p_{\rm d} = 3.03$, $p_{\rm a} = 2.8$ or (d)-(f) negative surface defect waveguide at $p_{\rm d} = 2.7$, $p_{\rm a} = 2.8$. Input peak power is (a) $0.2$ MW, (b) $1.3$ MW, (c) $2.7$ MW, (d) $0.2$ MW, (e) $1.3$ MW, and (f) $3.2$ MW.

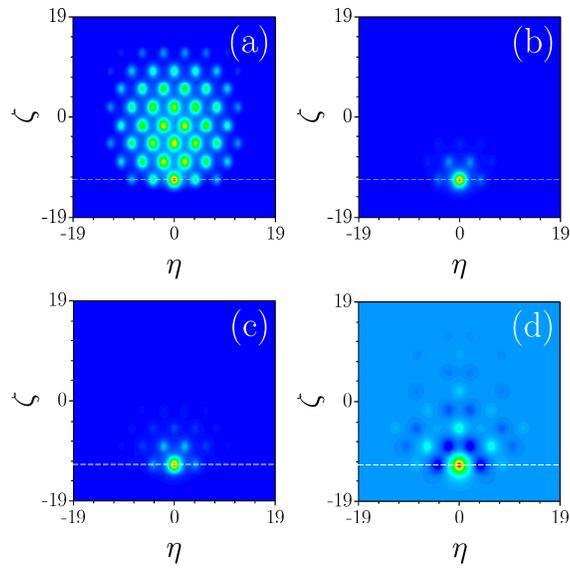

Figure 1. Field distributions for defect surface solitons at (a) $b = 0.4644$, $p_\mathrm{d} = 2.96$, (b) $b = 0.512$, $p_\mathrm{d} = 2.96$, (c) $b = 0.49$, $p_\mathrm{d} = 3.1$, (d) $b = 0.274$, $p_\mathrm{d} = 1.7$. The white dashed lines indicate the interface position.

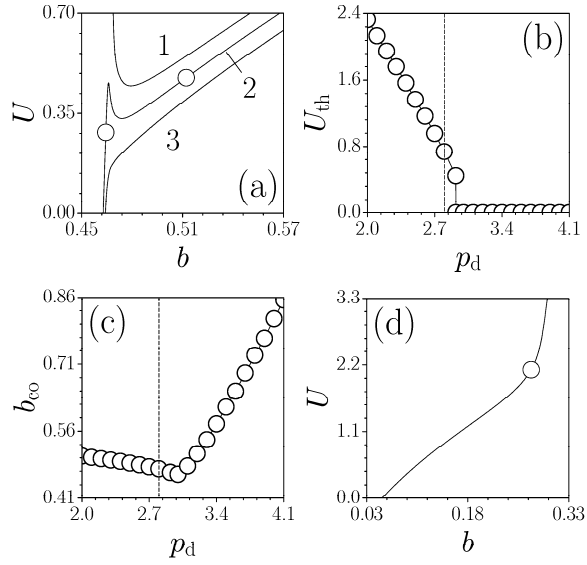

Figure 2. (a) Power versus propagation constant for $p_d = 2.92$ (curve 1), 2.96 (curve 2), and 3 (curve 3). The circles correspond to solitons in Figs. 1(a) and 1(b). Threshold power (b) and cutoff (c) versus $p_d$. Dashed lines indicate the point $p_d = 2.8$. (d) Power versus propagation constant for a surface mode existing due to Bragg guiding at $p_d = 1.7$. The circle corresponds to profile from Fig. 1(d).

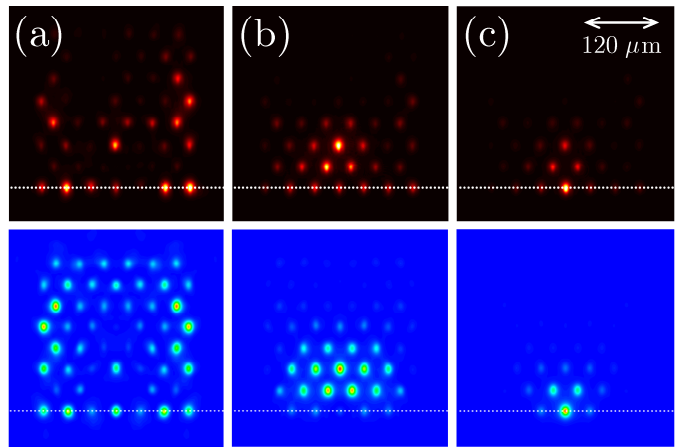

Figure 3. Comparison of the output intensity distributions for an excitation of a surface waveguide in array without defect at $p_\mathrm{d} = p_\mathrm{a} = 2.8$. Top row - experiment, bottom row - theory. The input peak power is (a) 0.18 MW, (b) 1.3 MW, and (c) 3.2 MW.

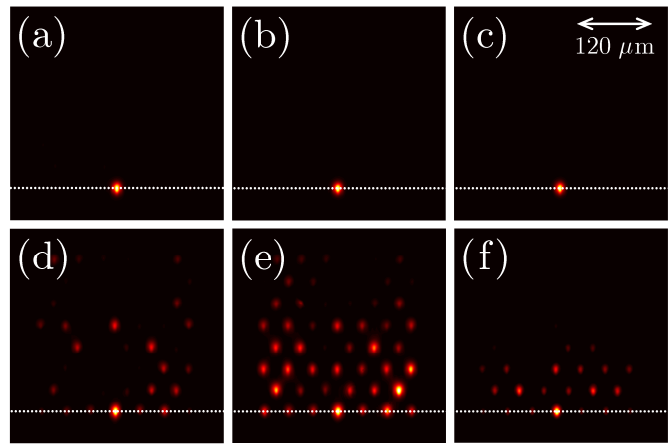

Figure 4. Output intensity distributions for an excitation of (a)-(c) positive surface defect waveguide at $p_\mathrm{d} = 3.03$, $p_\mathrm{a} = 2.8$ or (d)-(f) negative surface defect waveguide at $p_\mathrm{d} = 2.7$, $p_\mathrm{a} = 2.8$. Input peak power is (a) 0.2 MW, (b) 1.3 MW, (c) 2.7 MW, (d) 0.2 MW, (e) 1.3 MW, and (f) 3.2 MW.